\input harvmac

\overfullrule=0pt
\def\Title#1#2{\rightline{#1}\ifx\answ\bigans\nopagenumbers\pageno0\vskip1in
\else\pageno1\vskip.8in\fi \centerline{\titlefont #2}\vskip .5in}

\font\ticp=cmcsc10

\font\secfont=cmcsc10

%
%defs
%
\baselineskip=18pt plus 2pt minus 2pt

\def\ajou#1&#2(#3){\ \sl#1\bf#2\rm(19#3)}
%
%Caligraphic
\def\CH{{\cal H}}

\def\CF{{\cal F}}

\def\CL{{\cal L}}

%
%\def\eightbig#1{{\hbox{$\textfont0=\ninerm\textfont2=\niesy\left#1\vbox
% to6.5pt{}
%\right.\n@space$}}}
%\eightbig{N}

%Roman

%

%Greek

\def\g{\gamma}

\def\a{\alpha}
\def\b{\beta}
\def\d{\delta}

\def\p{\pi}
\def\e{\epsilon}

\def\l{{\lambda}}

%

%This paper

\def\TrH#1{ {\raise -.5em
                      \hbox{$\buildrel {\textstyle  {\rm Tr } }\over
{\scriptscriptstyle \CH _ {#1}}$}~}}

\def\IZ{\relax\ifmmode\mathchoice
{\hbox{\cmss Z\kern-.4em Z}}{\hbox{\cmss Z\kern-.4em Z}}
{\lower.9pt\hbox{\cmsss Z\kern-.4em Z}}
{\lower1.2pt\hbox{\cmsss Z\kern-.4em Z}}\else{\cmss Z\kern-.4em Z}\fi}
\def\IC{\relax\hbox{$\inbar\kern-.3em{\rm C}$}}
\def\IR{\relax{\rm I\kern-.18em R}}
\def\1{\relax 1 { \rm \kern-.35em I}}
\font\cmss=cmss10 \font\cmsss=cmss10 at 7pt

%

%Shortforms
\def\frac#1#2{{#1 \over #2}}
\def\ie{{\it i.e.}}

\def\p+{{\partial_+}}

\def\half{{1 \over 2}}

%

%refs

%

\Title{\vbox{\baselineskip12pt
\hbox{CALT-68-2038}
\hbox{\ticp doe research and}
\hbox{\ticp development report}
\hbox{}
\hbox{hep-th/9602030}
}}
{\vbox{\centerline {\bf AN ORIENTIFOLD OF TYPE-IIB THEORY ON K3} }}

\centerline{{\ticp
Atish Dabholkar\footnote{$^1$}{e-mail: atish@theory.caltech.edu}
and Jaemo Park\footnote{$^2$}{e-mail: jpk@theory.caltech.edu}
}}

\vskip.1in
\centerline{\it Lauritsen Laboratory of  High Energy Physics}
\centerline{\it California Institute of Technology}
\centerline{\it Pasadena, CA 91125, USA}

\vskip .1in

\bigskip
\centerline{ABSTRACT}
\medskip

A new orientifold of Type-IIB theory on $K3$ is constructed
that has $N=1$ supersymmetry in six dimensions. The orientifold
symmetry consists of a $Z_2$ involution of $K3$ combined with
orientation-reversal on the worldsheet.
The closed-string sector in the resulting theory
contains nine tensor multiplets and
twelve neutral hypermultiplets in addition to the gravity multiplet,
and is anomaly-free by itself.
The open-string sector contains only 5-branes and
gives rise to maximal gauge groups $SO(16)$
or $U(8)\times U(8)$ at different points in the moduli space.
Anomalies are canceled by a generalization of
the Green-Schwarz mechanism that involves more than one
tensor multiplets.

\bigskip

\bigskip
%\baselineskip=20pt plus 2pt minus 2pt
%\draft
\Date{February, 1996}

\vfill\eject

%References
\def\npb#1#2#3{{\sl Nucl. Phys.} {\bf B#1} (#2) #3}
\def\plb#1#2#3{{\sl Phys. Lett.} {\bf B#1} (#2) #3}

\def\mpl#1#2#3{{\sl Mod. Phys. Lett. }{\bf #1} (#2) #3}

%-------------------
% references
%-------------------
%

\lref\SaPe{
A. Sagnotti, in Cargese '87, ``Non-perturbative Quantum
Field Theory,'' ed. G. Mack et. al. (Pergamon Press, 1988) p. 521\semi
V. Periwal, unpublished.}

\lref\Hora{
P. Horava, \npb{327}{1989}{461}.}

\lref\DLP{
J. Dai, R. G. Leigh, and J. Polchinski,
\mpl{A4}{1989}{2073}\semi
R. G. Leigh, \mpl{A4}{1989}{2767}. }

\lref\ScSe{J. H. Schwarz and A. Sen, \plb{357}{1995}{323}, hep-th/9507027.}

\lref\Sen{A. Sen, ``M-Theory on $(K3 \times S^1 )/Z_2$,'' MRI-PHY/07/96,
hep-th/9602010.}

\lref\GiPo{E.~G.~Gimon and J.~Polchinski, ``Consistency Conditions
for Orientifolds and D-manifolds,'' hep-th/9601038.}

\lref\DMW{M.~J.~Duff, R.~Minasian, and E.~Witten,
``Evidence for Heterotic/Heterotic Duality,''
CTP-TAMU-54/95, hep-th/9601036.}

\lref\Polc{
J.~Polchinski, ``Dirichlet Branes and Ramond-Ramond
Charges,'' to appear in {\sl Phys. Rev. Lett.}, hep-th/9510017.}

\lref\toappear{work in progress.}

\lref\Erle{J.~Erler, ``Anomaly Cancellation in Six Dimensions,''
{\sl J. Math. Phys.}{\bf 35}(1988) 377.}

\lref\Stro{A. Strominger, \npb{451}{1995}{96}, hep-th/9504090\semi
B. Greene, D.~Morrison, and A. Strominger, \npb{451}{1995}{109}.}

\lref\GrSc{
M. B. Green and J. H. Schwarz, \plb{149}{1984}{117}\semi
\plb{151}{1985}{21}.}

\lref\PoCa{
J. Polchinski and Y. Cai, \npb{296}{1988}{91}.}

\lref\CLNY{
C. G. Callan, C. Lovelace, C. R. Nappi and S.A. Yost,
\npb{308}{1988}{221}.}

\lref\EGH{T. Eguchi, P.~B.~Gilkey, and A.~J.~Hanson,
{\sl Phys. Rept.} {\bf 66} (1980) 213.}

\lref\CLNYII{
C. G. Callan, C. Lovelace, C. R. Nappi and S.A. Yost,
\npb{293}{1987}{83}.}

\lref\GSWII{M. B. Green, J. H. Schwarz, and E. Witten,
{\it Superstring Theory},  {\rm Vol. II} ,
Cambridge University Press (1987).}

\lref\AGWi{L. Alvarez-Gaum\'e and E. Witten,
\npb{234}{1983}{269}.}

\lref\Schw{ J. Schwarz, ``Anomaly-Free Supersymmetric Models in Six
Dimensions,'' preprint  CALT-68-2030, hep-th/9512053.}

\lref\SagnI{A. Sagnotti, \plb{294}{1992}{196}.}

\lref\HoraII{P. Horava, \plb{231}{1989}{251}.}

\lref\Roma{L.~J.~Romans, \npb{276}{1986}{71}\semi
H.~Nishino and E.~Sezgin, \npb{278}{1986}{353}.}

\lref\Vafa{C.~Vafa, ``Evidence for F-theory,'' HUPT-96/A004,
hep-th/9602022.}

\lref\Witt{E.~Witten, ``Five-branes and M-Theory on an Orbifold,''
hep-th/9512219.}

\lref\WittII{E. Witten, ``Some Comments on String Dynamics,''
hep-th/9507121}

\lref\DaMu{K.~Dasgupta and S.~Mukhi, ``Orbifolds of M-Theory,''
hep-th/9512196.}

\lref\DSW{M.~Dine, N.~Seiberg, E.~Witten, \npb{289}{1987}{589}.}

\lref\ToSe{
P. Townsend, \plb{139}{1984}{283}\semi
N. Seiberg, \npb{303}{1988}{286}.}

\lref\SagnII{M. Bianchi and A. Sagnotti,
\plb{247}{1990}{517}; \npb{361}{1991}{519}\semi
A. Sagnotti, {\it Some Properties of Open-String
Theories,} preprint ROM2F-95/18, hep-th/9509080.}

%Unused references

\newsec{Introduction}

Theories of unoriented strings
can be viewed  as orientifolds
\refs{\SaPe, \Hora, \DLP} of oriented closed strings.
Orientifolds are a generalization of  orbifolds in which
the orbifold symmetry includes orientation reversal on the
worldsheet. For example, Type-I strings can be viewed as an
orientifold of Type-IIB strings.
It is obvious that the closed-string sector of unoriented
strings can be obtained by projecting the spectrum of
oriented strings onto states that are invariant under the
orientifold symmetry.
It is more difficult to see how and when the open string sector might arise,
and in particular how to obtain the Chan-Paton factors.
A proper understanding of this question has become possible only after
the remarkable recent work on D-branes\Polc.

A D-brane is a submanifold where strings are allowed to end
which corresponds to
open strings that satisfy mixed Dirichlet and Neumann boundary
conditions. In Type-II theories, D-branes represent
non-perturbative extended states that are charged
with respect to the R-R fields in the theory.
D-branes provide a geometric understanding of
how Chan-Paton factors arise:
a Chan-Paton label is simply the label of the D-brane that an open
string ends on.

One can now understand the open-string sector of an
orientifold as follows.
Orientifolding introduces unoriented surfaces
in the closed-string perturbation theory.
The unoriented surfaces such as the Klein bottle
can have tadpoles of R-R fields in the closed string tree
channel. The tadpoles can be canceled by including
the right number of D-branes that couple to these R-R fields.
This introduces the open string sector with
appropriate boundary conditions and Chan-Paton factors.

With this enhanced understanding of orientifolds,
one can contemplate more general constructions.
In this paper we construct a simple orientifold of
Type-IIB theory compactified on a $K3$ surface that
has $N=1$ supersymmetry  in six dimensions.
The orientifold symmetry group is $\{ 1, \Omega S \}$
where $S$ is a $Z_2$ involution of $K3$ and $\Omega$ is orientation
reversal on the worldsheet.
The resulting closed string sector contains the gravity multiplet, nine
tensor multiplets, and twelve neutral hypermultiplets. The maximal
gauge group arising from the open string sector is $SO(16)$ with
an adjoint hypermultiplet, or $U(8) \times U(8)$ with two
hypermultiplets that transform as $(8, 8)$.

There are a number of motivations for considering this example.
First, the requirement of anomaly cancellation in six dimensions
is fairly restrictive and provides useful constraints on
the construction of the worldsheet theory.
In fact, this work
was motivated in part by the observation \Schw\ that anomalies
cancel in a large class of supersymmetric
models in six dimensions. The orientifold that we
consider realizes one of these models as a string theory.
Second, we obtain
a massless spectrum that is markedly different from the
only known string compactification to six
dimensions with $N=1$ supersymmetry {\it viz.} the heterotic string
theory on $K3$, which has only one tensor multiplet. We thus have
a new compactification with a moduli space that apparently
is disconnected from the known compactifications.
Finally, this orientifold
is a useful practice case for various generalizations to different
dimensions using other orientifold groups \toappear.

The organization of the paper is as follows.
In section two we motivate the orientifold group from
considerations of anomaly cancellation and describe
the closed string sector.
The open string sector is discussed in section three.
Consistency requires inclusion of
$32$ Diriclet $5$-branes but {\it no}
$9$-branes, with
additional constraints on the Chan-Paton factors that
determine the gauge group and matter representations completely.

\newsec{Gravitational Anomalies and the Orientifold group}

The massless representations of the $N=1$ supersymmetry algebra
in $d=6$ are chiral;
consequently their coupling to gravity is potentially
anomalous.
We would like to see what constraints are placed on the massless
spectrum so that these anomalies cancel. We shall then use
this information to see how such a spectrum may follow from a string
compactification.

The massless states are labeled by the representations
of the little group in six dimensions which is $SO(4)=SU(2) \times SU(2)$.
The massless $N=1$ supermultiplets are then as follows.

1. The gravity multiplet:

\qquad a graviton (3, 3),
a gravitino 2(2, 3),
a self-dual two-form $(1, 3)$.

2. The vector multiplet:

\qquad a gauge boson  (2, 2),
a gaugino 2(1, 2).

3. The tensor multiplet:

\qquad an anti-self-dual two-form $(3, 1)$,
a fermion $2(2, 1)$, a scalar $(1, 1)$.

4. The hypermultiplet:

\qquad four scalars $4(1, 1)$,
a fermion $2(2, 1)$.

The gravitino and the gaugino are right-handed whereas the fermions
in the other two multiplets are left-handed. Up to overall normalization
the gravitational anomalies are given by \refs{\AGWi ,\GSWII}
\eqn\anomaly{\eqalign{
I_{3/2} &= -\frac{43}{288}(tr R^2)^2 + \frac{245}{360} tr R^4, \cr
I_{1/2} &= +\frac{1}{288}(tr R^2)^2 + \frac{1}{360} tr R^4, \cr
I_{A} &= -\frac{8}{288}(tr R^2)^2 + \frac{28}{360} tr R^4. \cr
}}
Here $I_{3/2}$, $I_{1/2}$, and $I_A$ refer to the anomalies for
the gravitino, a right-handed fermion,
and a  self-dual two-form $(1, 3)$ respectively.

Consider $n_V$ vector multiplets, $n_H$ hypermultiplets and
$n_T + 1$ tensor multiplets. Then
the $(tr R^4)$ term cancels if the following condition is
satisfied:
\eqn\condition{n_H - n_V = 244 - 29 n_T.}
The $(tr R^2)^2$ term is in general nonzero, and needs to be canceled
by the Green-Scwarz mechanism \GrSc. There are many solutions
of \condition. We would now like to see which can be realized
as a string theory.

There are not many possibilities for string vacua
with $N=1$ supersymmetry in six dimensions.
For the heterotic string,
we must compactify on a $K3$ to obtain $N=1$ supersymmetry.
This leads to $n_T=0$ and $n_H = n_V +244$.
For Type-II strings, usual Calabi-Yau compactification on a $K3$ leads
to $N=2$ supersymmetry.
One way to reduce supersymmetry further is to take an orientifold  so that
only one combination of the left-moving and the right-moving supercharges
that is preserved by the orientation-reversal survives. By considering
different orientifold groups one may obtain different spectra, and in
particular different number of tensor multiplets.

The model that we consider in this paper has $n_T =8$
and $n_H - n_V =12$ which clearly satisfies \condition.
The special thing that happens with this matter content is that the
entire anomaly polynomial including  the $(tr R^2)^2$ term vanishes.
We thus have anomaly cancellation without the need for the Green-Schwarz
mechanism, analogous to
what happens in the Type-IIB theory
in ten dimensions \AGWi, or in the chiral $N=2$ theory
obtained by compactifying  Type-IIB theory on $K3$ \ToSe.

If we wish to obtain a large number of tensor multiplets, a
natural starting point for orientifolding is
the Type-IIB theory compactified on $K3$, which has
$21$ ($N=2$) tensor multiplets in the massless spectrum
in addition to the gravity multiplet.
The gravity multiplet contains
$5$ self-dual two-forms whereas the tensor multiplets contain
one anti-self-dual two-form each. Let us recall how these two-forms
arise. In ten dimensions
the Type-IIB theory contains a two-form $B^1_{MN}$
from the R-R sector,
a two-form $B^2_{MN}$ from the NS-NS sector and a four-form
$A_{MNPQ}$ from the R-R sector with self-dual field strength.
Zero modes of these fields
correspond to harmonic forms on $K3$ and give rise to
massless fields in six dimensions \GSWII .
The nonzero Betti numbers for $K3$ are
$b_0 = b_4 =1$, $b^{+}_2 =3$, and $b^{-}_2 =19$ where
$b^+_2$ are the self-dual two-forms and $b^-_2$ are the
anti-self-dual two-forms. {}From the two $B_{MN}$ fields
we get $b_0$ two-forms each, which means altogether $2$ self-dual and
$2$ anti-self-dual two-forms. Similarly, from the zero modes of the
$A_{MNPQ}$ we get $3$ self-dual and $19$ anti-self-dual
two-forms in six dimensions after imposing self-duality of
field strength in ten dimensions.

The orientifold group can now be deduced as follows. In order
to obtain $N=1$ supersymmetry
we need an orientation reversal $\Omega$ which takes
$\sigma$ to $\pi -\sigma$. A projection $(1 + \Omega )/2$ alone
would give us the spectrum identical to the closed-string sector
of Type-I theory on $K3$, eliminating
$A_{MNPQ}$ and $B^2_{MN}$ completely from the spectrum.
Now consider a $Z_2$ involution $S$ of $K3$ such that
eight anti-self-dual harmonic forms are odd under $S$ and
all other $16$ forms are even. It is clear that under the projection
$(1 + \Omega S)/2$, eight zero-modes of $A_{MNPQ}$
will now survive, giving us $8$ anti-self-dual
two-forms. Moreover, we shall also get eight scalars
from the zero modes of $B^2_{MN}$ so that we have the complete
bosonic content of eight tensor multiplets. We still have
one zero mode of $B_{MN}^1$ giving one self-dual
and one anti-self-dual two-form. The self-dual two-form
is needed for the gravity multiplet; the anti-self-dual
two-form combines with the zero mode of the dilaton to
form an additional tensor multiplet. Altogether, we obtain the nine
tensor multiplets that we were after.

Let us see if we get the rest of the spectrum right.
There are no vector multiplets because
there are no odd cycles on $K3$, and starting
with even forms and the metric in ten dimensions
we can never get a one-form as a zero mode.
The scalars arise from zero modes
of the metric tensor and the $B^1_{MN}$ field that are
invariant under $\Omega S$. Their zero modes can be found
from the Dolbeault cohomology of $K3$ \GSWII,  so we need to
know which $(p, q)$ forms are left invariant by $S$.
The main point for our purpose will be that
the eight two-forms that are eliminated by $S$ are $(1, 1)$
forms
\foot{For a smooth $K3$  defined by a
quartic polynomial in ${CP}^3$, it is easy to
construct an example of the involution $S$ and
verify this assertion \ScSe.}.
We are thus left with $12$ $(1,1)$ forms
and $1$ each of
$(0,2)$, $(2, 0)$, $(0, 0)$, $(2, 2)$ forms.
The zero modes of $g_{MN}$
give $34$ scalars \GSWII. The number of zero modes of
$B^1_{MN}$ equals the number of harmonic two-forms which is
$14$. Altogether we have $48$ scalars which make up $12$
hypermultiplets.
This construction ensures
that the closed-string sector is anomaly free. We also
get a constraint in the open-string sector
that the number of vector multiplets must equal
the number of hypermultiplets for canceling gravitational anomalies.

To proceed further we need to know the spectrum in
the open-string sector and check that all tadpoles
vanish. For computing the tadpoles
we need a realization of
the $K3$ as an explicit worldsheet conformal field
theory. Furthermore, we need to know how the involution $S$
acts in this conformal field theory. This can be easily
done for a particular $K3$ represented as a $T^4/Z_2$ orbifold.
Let $(z_1, z_2)$ be complex coordinates on the torus
$T^4$ defined by periodic identifications
$z_1 \sim z_1 + 1$, $z_1 \sim z_1 + i $,
and similarly for $z_2$.
The two $Z_2$ transformations
of interest are generated by
\eqn\discrete{\eqalign{
R: &\quad (z_1, z_2) \rightarrow (-z_1, -z_2) \cr
S: &\quad (z_1, z_2) \rightarrow (-z_1 + \half, -z_2 + \half ). \cr
}}
That $S$ is the desired symmetry can be seen as follows.
The $K3$ orbifold is obtained by dividing the torus by
$Z^R_2 \equiv \{1, R\}$.
The Type-IIB theory on this orbifold has $5$ self-dual and $5$
anti-self-dual two-forms coming from the untwisted sector.
In the twisted sector, there are $16$ anti-self-dual forms from the
$16$ fixed points of $R$. Notice that $S$ is the same as $R$ acting
on shifted coordinates $(z_1 -\frac{1}{4}, z_2 -\frac{1}{4})$.
Now, $S$ leaves all forms in the untwisted
sector invariant, but takes $8$ fixed points of $R$ into
the other $8$. Thus of the anti-self-dual two-forms coming from the
twisted sector, $8$ are even under $S$,
and $8$ are odd. This is precisely the structure we wanted.
Note that $S$ has $16$ fixed points on the torus,
but on the orbifold they are
identified under $R$ leaving only $8$ as required by
the Lefschetz fixed-point theorem \EGH.

\newsec{Open String Sector}

\subsec{Tadpoles}

Tree-channel tadpoles can be evaluated by factorizing the partition
function in the loop channel.
For closed strings,  the one-loop amplitude for the orientifold
is obtained by projecting onto the closed string states of the Type-IIB
theory on $K3$ that are invariant under the symmetry $\Omega S$.
The partition function now receives a contribution
from the Klein bottle in addition to the torus.
The torus has no closed-string tree channel and is modular invariant
by itself, so we need to consider only the Klein bottle.
To determine the open string sector we first
require closure of operator product expansion so that
the S-matrix factorizes properly. This implies that we
can consistently add only $5$-branes and $9$-branes \GiPo.
We then have $55, 99, 59, 95$ sectors for open strings
from strings that begin and end on the two kinds of branes.
The one-loop partition function is given by the cylinder and
the M\"obius strip diagram.

In this section we shall follow the general framework
of Gimon and Polchinski \GiPo quite closely.
The total projection that we wish to perform
is $(\frac{1+R}{2})(\frac{1+\Omega S}{2})$.
The orientifold group $G$ is
$\{1, R, \Omega S, \Omega RS\}$ which we can write as
$G=G_1 +\Omega G_2$ with $G_1 = \{1, R\}$
and $G_2=\{S, RS\}$.
An open string can begin on a D-brane labeled by $i$ and end
on one labeled by $j$. The label of the D-brane is the
Chan-Paton factor at each end.
Let us denote a  general state in the open string sector
by $|\psi, ij \rangle$. An element of $G_1$ then acts
on this state as
\eqn\gone{
g:\qquad |\psi,ij\rangle \ \rightarrow
(\gamma_g )_{ii'}|g\cdot\psi,i'j'\rangle
(\gamma_g^{-1})_{j'j},}
for some unitary matrix $\gamma_g$ corresponding to $g$.
Similarly, an element of $\Omega G_2$ acts as
\eqn\gtwo{
\Omega h:\qquad |\psi,ij\rangle \ \rightarrow
(\gamma_{\Omega h})_{ii'}|\Omega h\cdot\psi,j'i'\rangle
(\gamma_{\Omega h}^{-1})_{j'j}.}

The relevant partition sums for the Klein bottle,
the M\"obius strip, and the cylinder  are respectively
$\int_0^{\infty}dt/2t$ times
\eqn\traces{\eqalign{
{\rm KB:} &\quad {\rm Tr}_{\rm NSNS + RR}^{\rm U+T}
\left\{ \frac{\Omega S}{2}\,\frac{1+R}{2}\,\frac{1+(-1)^{F}}{2}
e^{-2\pi t(L_0 + \tilde L_0)}\right\} \cr
{\rm MS:} &\quad {\rm Tr}_{\rm NS-R}^{99+55}
\left\{ \frac{\Omega S}{2}\,\frac{1+R}{2}\,\frac{1+(-1)^{F}}{2}
e^{-2\pi t L_0}\right\} \cr
{\rm C:} &\quad {\rm Tr}_{\rm NS-R}^{99+95+59+55}
\left\{ \frac{1}{2}\,\frac{1+R}{2}\,\frac{1+(-1)^{F}}{2}
e^{-2\pi t L_0}\right\} .\cr }}
Here $F$ is the worldsheet fermion number,
and as usual $\frac{1+(-1)^{F}}{2}$ performs the GSO projection.
The Klein bottle includes contributions both from the
untwisted sector and the sector twisted by $R$ of the original
orbifold.

For evaluating the traces we need to know the action of
various operators on the oscillator modes and the zero modes of the fields.
Let us take $X^m, m=6,7,8,9$ to be the coordinates of the torus
so that $2\pi r z_1 = X^6+iX^7$ and $2\pi r z_2 = X^8 +i X^9$,
where the radius $r$ defines the overall size of the torus.
Let $X^i, i=1, 2, 3, 4$
be the transverse coordinates in the six-dimensional Minkowski space.
Let $\psi^m$ and $\psi^i$ be the corresponding fermionic coordinates
of the NSR string.
The action of $R$ on oscillator modes is obvious.
For the ground states
$|p_m, L^m\rangle$
without oscillations, but with quantized
momentum $p_m\equiv k_m/R$ in the compact
direction and winding $L^m \equiv X^m (2\pi ) - X^m(0)$,
R has the action
\eqn\raction{
R |p_m ,  L^m\rangle = |-p_m , -L^m\rangle .}
Note that $S$ is $U(\frac{r}{4}) R U^\dagger (\frac{r}{4})$
where $U(\frac{r}{4})$ performs translation along both $X^6$ and
$X^8$ by $r/4$. Therefore, $S$ has the same action on the oscillators as $R$
but for the ground states there is a crucial difference of phase
\eqn\saction{
S |p_m ,  L^m\rangle = (-1)^{k_6}(-1)^{k_8} |-p_m , -L^m\rangle.}
The action of $\Omega$ depends on the sectors;
$\Omega$ takes a field $\phi (\sigma )$ to $\phi (\pi - \sigma )$
and has obvious action on the modes.

The traces can be readily evaluated.
Following \GiPo\ we define
\eqn\jacobi{\eqalign{
f_{1}(q) = q^{1/12} \prod_{n=1}^\infty \left(1-q^{2n}\right),\qquad
& f_{2}(q) = q^{1/12} \sqrt{2}\,\prod_{n=1}^\infty \left(1+q^{2n}\right)\cr
f_{3}(q) = q^{-1/24} \prod_{n=1}^\infty \left(1+q^{2n-1}\right),\qquad
& f_{4}(q) = q^{-1/24}\prod_{n=1}^\infty \left(1-q^{2n-1}\right),\cr}}
which satisfy the Jacobi identity
\eqn\ident{f_{3}^{8}(q) = f_{2}^{8}(q)+f_{4}^{8}(q)}
and have the modular transformations
\eqn\modular{
f_{1}(e^{-{\pi}/{s}}) = \sqrt{s}\,f_{1}(e^{-{\pi} s}),\quad
f_{3}(e^{-{\pi}/{s}}) = f_{3}(e^{-\pi s}),\quad
f_{2}(e^{-{\pi}/{s}}) = f_{4}(e^{-\pi s}).
}
The relevant amplitudes are then given by
$
(1-1)\frac{v_6}{128} \int_0^\infty \frac{dt}{t^4}
$
times
\eqn\amplitude{\eqalign{
{\rm KB:} &\quad
8\frac{f_{4}^{8}(e^{-2\pi t})}{f_{1}^{8}(e^{-2\pi t})}\left\{
\left( \sum_{n=-\infty}^\infty (-1)^n e^{-\pi t n^{2}/\rho} \right)^{2}
\left( \sum_{n=-\infty}^\infty  e^{-\pi t n^{2}/\rho} \right)^{2}
+ \left( \sum_{w=-\infty}^\infty e^{-\pi t \rho w^{2}} \right)^{4}\right\}\cr
{\rm MS:} &\quad - \frac{f_{2}^{8}(e^{-2\pi t})f_{4}^{8}(e^{-2\pi t})}
{f_{1}^{8}(e^{-2\pi t})f_{3}^{8}(e^{-2\pi t})}\left\{
\quad {\rm Tr}(\gamma_{\Omega S,5}^{-1}\gamma_{\Omega S,5}^T)
\left(\sum_{w=-\infty}^\infty
e^{-2\pi t\rho w^{2}} \right)^{4}\right.\cr
&\left. \quad +{\rm Tr}(\gamma_{\Omega RS,9}^{-1}\gamma_{\Omega RS,9}^T)
\left(\sum_{n=-\infty}^\infty (-1)^n e^{- 2\pi t n^{2}/\rho} \right)^{2}
\left(\sum_{n=-\infty}^\infty e^{- 2\pi t n^{2}/\rho} \right)^{2}\right\}\cr
{\rm C:}&\quad \frac{f_{4}^{8}(e^{-\pi t})}{f_{1}^{8}(e^{-\pi t})}
\left\{ ({\rm Tr}(\gamma_{1,9}))^2
\left(\sum_{n=-\infty}^\infty e^{- 2\pi t n^{2}/\rho} \right)^{4}
 \right.\cr
& \left.\qquad + \sum_{i,j \in 5} (\gamma_{1,5})_{ii} (\gamma_{1,5})_{jj}
\prod_{m=6}^9 \sum_{w=-\infty}^\infty
e^{-t (2\pi wr +X^m_i - X^m_j)^{2}/2\pi\alpha'} \right\}\cr
&- 2  \frac{f_{2}^{4}(e^{-\pi t})f_{4}^{4}(e^{-\pi t})}
{f_{1}^{4}(e^{-\pi t})f_{3}^{4}(e^{-\pi t})} {\rm Tr}(\gamma_{R,5})
{\rm Tr}(\gamma_{R,9}) \cr
&+ 4 \frac{f_{3}^{4}(e^{-\pi t})f_{4}^{4}(e^{-\pi t})}
{f_{1}^{4}(e^{-\pi t})f_{2}^{4}(e^{-\pi t})}
\left\{ ({\rm Tr}(\gamma_{R,9}))^2 + \sum_{I=1}^{16}
({\rm Tr}(\gamma_{R,I}))^2 \right\}.\cr
}}
We have defined $v_6 = V_6 /(4\pi^2 \alpha')^3$ where $V_6$ is the
(regulated) volume of the non-compact dimensions,
and $\rho = r^2/\alpha'$.
For the cylinder amplitude, as in \GiPo, the sum $i, j$ comes from
strings that begin and end at 5-branes $i$ and $j$ with arbitrary
windings; the sum $I$ is over 5-branes placed at the fixed points
of $R$. Note that for the Klein bottle and the M\"obius
strip diagrams, in evaluating ${\rm Tr}( \Omega RS )$ or
${\rm Tr} (\Omega S )$,
the sum over momenta contains a crucial factor of $(-1)^n$ for
the $6$ and $8$ directions,
but no such factor for the $7$ and $9$ directions.

To factorize in tree channel we use the modular transformations
\modular\
and the Poisson resummation formula
\eqn\poisson{
\sum_{n=-\infty}^{\infty} e^{-\pi (n-b)^2/a} =
\sqrt{a} \sum_{s=-\infty}^{\infty} e^{-\pi a s^2 + 2\pi i s b}.}
An important fact for our purpose
will be that
\eqn\poissontwo{
\sum_{n=-\infty}^{\infty}(-1)^n e^{-\pi t n^2/\rho} =
\sqrt{\rho/t}
\sum_{s=-\infty}^{\infty} e^{-\pi \rho (s-\half)^2/t}.}

Tadpoles correspond to long tubes$(t\rightarrow 0)$ in the tree channel.
In this limit
it is easy to see that the total amplitude is proportional
to $(1-1) \int_0^\infty dl$
times
\eqn\tadpole{\eqalign{
&\frac{v_6 v_4}{16} \left\{ ({\rm Tr}(\gamma_{1,9}))^2
\right\}
+\frac{v_6}{16 v_4} \left\{ 32^2 - 64 {\rm
Tr}(\gamma_{\Omega S,5}^{-1}\gamma_{\Omega S,5}^T) + ({\rm
Tr}(\gamma_{1,5}))^2
\right\}\cr
+& \frac{v_6}{64} \sum_{I=1}^{16} \left({\rm
Tr}(\gamma_{R,9}) - 4{\rm Tr}(\gamma_{R,I}) \right)^2. \cr}}
Here $l$ is the length of the tube, which is inversely
proportinal to the loop modulus $t$;
$v_4 =\rho^2 = V_4/(4\pi^2 \alpha')^2$ with
$V_4$ the volume of the internal torus before orbifolding.

The $(1-1)$ above represents the contributions of NSNS and RR exchange
respectively,
which must vanish separately for consistency \refs{\PoCa, \CLNY}.
Using these requirements we determine the spectrum in the next
section.

\subsec{Gauge Group and Spectrum}

We see from \tadpole\ that to cancel the tadpole proportional
to $v_6 v_4$ corresponding to the 10-form exchange, we must
have ${\rm Tr} (\gamma_{1,9}) =0$. Now ${\rm Tr} (\gamma_{1,9})$
equals the number $n_9$ of 9-branes, so we conclude that
there are no 9-branes. We are left with only the $55$ sector
so from now on we drop the subscript $5$ for the $\gamma$
matrices. Vanishing of the term proportional to $v_6 /v_4$
correponding to the exchange of untwisted 6-forms gives
\eqn\solution{
n_5=32, \qquad \gamma_{\Omega S} = \gamma_{\Omega S}^T.}
Finally, vanishing of the term proportional to $v_6$ corresponding
to the exchange of twisted sector 6-forms gives
${\rm Tr}(\gamma_{R,I}) =0$.
By a unitary change of basis $\gamma_{\Omega S} \rightarrow
U \gamma_{\Omega S} U^T$
we can take
\eqn\omegas{
\gamma_{\Omega S} = {\bf 1}.}
We have additional constraints on the algebra of the $\g$
matrices so that we obtain
a representation of the orientifold group in the Hilbert space:
\eqn\rep{\eqalign{
\gamma_{\Omega RS} =&\gamma_{\Omega S}\,\gamma_{R} \cr
(\gamma_R )^2 =& {\bf 1}\cr
\gamma_{\Omega RS}^T=& \pm \gamma_{\Omega RS}.\cr
}}
We have the choice of
taking $\gamma_{\Omega RS}$ either symmetric or antisymmetric,
but it turns out that both choices lead to the same spectrum.

Let us now discuss the massless bosonic spectrum coming from
the $NS$ sector. The states
\eqn\vecotrs{
\psi_{-1/2}^\mu|0,ij\rangle \lambda_{ji}, \quad \mu = 1, 2, 3, 4,}
belong to the vector multiplets whereas the states
\eqn\scalars{\psi_{-1/2}^m|0,ij\rangle \lambda_{ji}, \quad  m= 6, 7, 8, 9,}
belong to the hypermultiplets.
We have to keep only the states that are
invariant under $R$ and $\Omega S$; this constrains the
possible forms of the Chan-Paton wave functions $\lambda_{ij}$.

The conditions for the Chan-Paton factors depend
crucially on where the 5-branes are placed.
There are a number of ways one can distribute the $32$ 5-branes
to obtain various gauge groups. We discuss only two distinct configurations
that lead to maximal symmetry.

1. The first choice  is to take $16$ five-branes to lie at
a fixed point $x$ of $S$ and the remaining $16$ to lie at
the image of $x$ under $R$. In this case, the projection
under $R$ simply relates the states at $x$ to those at $Rx$
and leads to no additional constraints on $\lambda$.
$\Omega S = +1$ implies
\eqn\chanpatone{
\lambda = - \gamma_{\Omega S} \lambda^T \gamma_{\Omega S}^{-1}}
for both scalars and vectors. This can be seen as follows.
$\psi^m$ satisfy Dirichlet boundary conditions on both ends
and have the same mode expansion as $\psi^{\mu}$ which
satisfy Neumann boundary conditions. Now $\psi^\mu_{-\half}$
is odd under $\Omega$ as
in Type-I theory in ten dimensions. But $\psi^m_{-\half}$
is even because of the additional phase due to the Dirichlet
boundary condition. Moreover, under $S$, $\psi^m$ is odd
and $\psi^{\mu}$ is even.
Using \omegas\ we conclude that $\l = -\l^T$, obtaining
an adjoint representation of $SO(16)$ for both vectors
and scalars, and the corresponding supermultiplets.

2. We can place $16$ five-branes at a fixed point $y$ of $R$
and $16$ at the image of $y$ under $S$. This time we only need
to impose the condition $R=+1$ on the states.
For the matrix $\g_R$ we had two choices. Let us first choose
$\g_{\Omega RS}$ to be symmetric. Then from \rep,
$\g_R$ is also a symmetric
matrix that squares to one and is traceless.
In transforming $\g_{\Omega S}$ to identity we
already made a unitary change of basis, but we can still
make an orthogonal change of basis to put
$\g_R$ in the form
\eqn\gammar{\g_R = \left(\matrix{{\bf 1}&0\cr
                0&-{\bf 1}\cr}\right).}
Now $R=1$ implies
$$ \l = \g_R \l \g_R^{-1}$$
for vectors and
$$ \l = - \g_R \l \g_R^{-1}$$
for scalars.
The condition for vectors means that we have
a subgroup of $U(16)$ that commutes with
$\g_R$ \ie, $U(8) \times U(8)$.
The condition for scalars means that they transform
as $(8, {\bar 8})$ and $({\bar 8}, 8)$ under the
$U(8) \times U(8)$. Another way to see this is to note
that the Chan-Paton label transforms
as $(1, 8)+(8, 1)$ at one end and as the complex
conjugate at the other. The projection
keeps $(8\times{\bar 8}, 1) +(1, 8\times{\bar 8})$ for the vectors,
and $(8, {\bar 8})$ and
the complex conjugate for the scalars.
If we chose $\g_{\Omega RS}$ antisymmetric, we would get
$\g_R =  \left(\matrix{0&-i{\bf 1}\cr
                i{\bf 1}&0\cr}\right)$ instead
of \gammar, but the identical spectrum.

Notice that the rank of the gauge group is different in
the two cases which correspond to two branches
of the moduli spaces that are connected.
With the group $SO(16)$ we have adjoint matter, so we cannot change
the rank. We can break it to a $U(8)$ or all the way to $U(1)^8$.
For the $U(8)\times U(8)$, the condensation of charged
hypermultiplets can change the rank and we can also break it
to the diagonal $U(8)$, for example. The two branches
are thus connected.

The symmetry breaking can be seen geometrically.
If we place a 5-brane away from the fixed points of
$R$ and $S$, then we need three more 5-branes at
the image points. We can thus divide the $32$ branes
in four copies of $8$.
In this case, there will be no restrictions on the Chan-Paton
matrices at a given point, except that they are hermitian.
If all branes are placed at generic
points and their images, we get $U(1)^8$.
When they coincide at a point other than the fixed points,
we get $U(8)$ with an adjoint hypermultiplet.

\subsec{Anomaly Cancellation}

The number of vector multiplets equals the number of hypermultiplets
at all points of the moduli space discussed in the previous
subsection,
so the gravitational anomalies cancel.
In fact, at a generic point in the moduli space where
the symmetry is $U(1)^8$, or also when it is $SO(16)$, the
entire anomaly vanishes. These theories are thus
anomaly-free without the need for the Green-Schwarz mechanism.

Anomaly cancellation is more subtle when the gauge group is
$U(8) \times U(8)$.
We can factorize the group
as $SU(8)\times SU(8)\times U(1)\times U(1)$. The states
are neutral under the diagonal $U(1)$.
So we need to consider only $SU(8)_1 \times SU(8)_2 \times U(1)$
under which the hypermultiplets transform as
$(8, {\bar 8})_+$ and $({\bar 8}, 8)_-$, where the
subscript denotes the $U(1)$ charge.
Let us denote the field strengths as $F_1, F_2$, and $f$
respectively.

The $U(1)$ factor is at first sight troublesome.
The anomaly involving this factor has terms that
are of the form $f ( d_1 {\rm tr} F_1^3 + d_2 {\rm tr} F_2^3 )$
where $d_1, d_2$ are constants.
Such terms would seem
problematic because they do not have the usual
factorized form $ f^2 {\rm tr} F^2$.
However, these can be canceled by a local
counterterm of the form $\int b {\Tr F^3}$
for some scalar $b$ that has
inhomogeneous gauge transformations.
Let $a$ be the gauge potential, $da = f$.
Under the gauge transformation $ \d a = d\e$,
$b$ must have the inhomogeneous transformation $\d b = \e$
to cancel the anomaly. The gauge invariant combination is
$A = db -a$ which is nothing but the gauge-invariant form
of the massive gauge boson associated with $a$.
Now the kinetic term for $b$ is of the
form $A^2$ which can be viewed as the mass term for
the massive gauge field $A$.

One is familiar with an analogous situation in four dimensions \DSW.
The scalar $b$ is very similar to the axion in four dimensions
which is the Goldstone boson of a global Peccei-Quinn symmetry.
The fermionic current for the Peccei-Quinn symmetry is anomalous, but
so is the axion current. Now, if we gauge this symmetry,
then naively we would find that the gauge coupling to the fermions
is anomalous. However, one can always define a linear combination of the
fermionic current and the axionic current which is anomaly-free.
The axion then is the would-be Goldstone boson associated
with this anomaly-free current.
The corresponding gauge-boson becomes massive after eating the
axion.

Because the $U(1)$ gauge boson will always be massive, we shall
discuss only the remaining factors $SU(8)_1 \times SU(8)_2$.
Let us denote the field strengths for the two groups by
$F_1$ and $F_2$ respectively, and define
$\CF_{\a} \equiv
{\rm tr} F_{\a}^2, \a =1,2$.
The anomaly polynomial is then of the form
\eqn\sixanomaly{
X = \CF_1^2 +\CF_2^2 -2\CF_1\CF_2.}
To cancel this anomaly one needs a
a generalization of the Green-Schwarz mechanism
proposed by Sagnotti \SagnI\  which we now review briefly.

If we have $n$ tensor multiplets, then there is
a natural $SO(1, n)$ symmetry in the low-energy
supergravity action \Roma. Altogether there are $n+1$ tensors
$H_r, r=0,...,n$ that transform as a vector of $SO(1, n)$;
the time-like component is self-dual whereas the spacelike
components are anti-self-dual.
The scalars coming
from the tensor multiplets parametrize the coset
space $SO(1, n)/SO(n)$.
We take $\eta_{rs}$ to be the Minkowski metric
with signature $(1, n)$.
Let $v$ be the time-like vector,
$v\cdot v = 1$,
so that $v\cdot H$ is self-dual.
The scalar product is with respect to
the metric $\eta$: for example, $v\cdot H \equiv v_r H_s \eta_{rs}$.
Now consider the case when the gauge group has $m$
nonabelian factors with field strengths
$F_{\a}, \a = 1,...,m$, and denote
${\rm tr} F_{\a}^2$ by $\CF_{\a}$. In this case,
anomaly cancellation can be achieved by a generalization
of the Green-Schwarz mechanism if the anomaly polynomial is
of the general form
\eqn\anopol{
X = -\sum_{\a\b}(c_{\a}\cdot c_{\b})\CF_{\a}\CF_{\b},}
where $c_{\a}, \a = 1,...,m$ are constant
vectors of $SO(1, n)$.
It is clear that the anomaly associated with $X$ can be canceled
by a local counterterm of the form
\eqn\counter{
\Delta \CL = \sum_{\a} \CF_{\a} (c_{\a} \cdot B),}
provided the fields $B_r$ have appropriate gauge transformations.
If $\omega_{\a}$ are the Chern-Simons three-forms
for the various gauge groups and
$\d \omega_{\a} = d\omega^1_{\a}$, then the required gauge
transformations are $\d B_r =  c_{\a r} \omega^1_{\a}$.
The modified gauge-invariant field strengths $H_r$
are then given by
\eqn\modfield{
H_r = dB_r - c_{\a r}\omega_{\a}.}
An important fact that follows from supersymmetry is
that the coefficients $c_{\a r}$ that enter into \modfield\ and
the modified Bianchi identity are related to
the kinetic term for the gauge field $F_{\a}$, which is given by
$v\cdot c_{\a}$ \SagnI. Given an anomaly polynomial,
the vectors $c_{\a}$ must be chosen such that the kinetic terms
for all gauge fields are positive-definite.

In our case, the gauge group has only two factors, \ie, $m=2$.
We have ten tensors $(n=9)$, but it turns out that only three tensors
are involved in the anomaly cancellation. This is because
when all branes are localized at a given fixed point
of $R$ (and its image under $S$), the tensors
coming from the twisted sectors localized at
other fixed points that are far away, cannot be relevant.
Therefore we restrict ourselves to a three dimensional
subspace taking $n=2$. We have one self-dual
and one anti-self-dual tensor from the untwisted sector,
and one anti-self-dual tensor from the twisted sector.

For simplicity, let us pick a special
point in the tensor-multiplet moduli space
so that $v = (\cosh{\phi}, \sinh{\phi}, 0)$.
The anomaly polynomial \sixanomaly\ can be
written in the form \anopol\ by choosing
$c_1 = (1, 1, 1)$ and
$c_2 = (1, 1, -1)$.
There is some freedom in choosing these vectors
because of the $SO(1, n)$ symmetry and the freedom in
choosing the signs of the tensor fields.
With the above choice
the field $\phi$ can be identified with the dilaton so that the
coefficient of the gauge kinetic term, which comes from the disk diagram,
goes as $e^{-\phi}$.
Moreover, the kinetic
terms are positive-definite for both the gauge groups
because $v\cdot c_1$ and $v\cdot c_2$ are both
positive-definite.
Thus, the anomalies
can be canceled by the generalized Green-Schwarz mechanism
explained in the preceding paragraphs.

Worldsheet considerations are consistent with this
spacetime reasoning. To obtain a counter-term
like \counter\ we would require a coupling
of the kind $B_2 ( \CF_1 -\CF_2 )$, where $B_2$
is the tensor coming from the twisted sector.
Such a term can be obtained by computing
a disk diagram with two vertex operators for
the gauge bosons on the boundary of the disk,
and the vertex operator for the tensor
at the center of the disk. The vertex operator
at the center introduces a branch-cut corresponding
to a twist by $R$. The twist acts on the Chan-Paton
indices by the matrix $\gamma_R$ which is
$+{\bf 1}$ for $\CF_1$ but $- {\bf 1}$
for $\CF_2$. This is in accordance with
the relative minus sign between the third
components of the two vectors $c_1$ and $c_2$.
By contrast, the vertex operators for
the two tensors $B_0$ and $B_1$ coming from the untwisted sector
of the orbifold introduce no branch cuts.
These tensors therefore have identical couplings to
the two gauge groups; correpondingly,
$c_1$ and $c_2$ are identical
in the $0, 1$ subspace.

We have not worked out the detailed couplings from a
worldsheet calculation, but
our tadpole calculation assures us that
anomaly must cancel in this way.
If gauge invariance were anomalous, then
the longitudinal mode of the gauge boson would
not  decouple.  This would lead to a tadpole, but
we have already made certain that there are no tadpoles.

So far we have chosen to work at a special point in the moduli space,
where one could ensure that
the kinetic terms for both gauge groups
are positive-definite. However, as we move around the tensor-multiplet
moduli space,
we eventually come across a boundary where
the kinetic term for one of the gauge fields changes sign, and
is no longer positive-definite.
For example, we can take a more general form for the vector $v$,
$v=(\cosh{\phi}, \sinh{\phi} \cos{\psi}, \sinh{\phi} \sin{\psi})$
where $\phi$ and $\psi$ are the moduli. It is easy to see
that there is a range of values for
$\phi$ and $\psi$ where either $v\cdot c_1$ or $v\cdot c_2$
is negative.
This phenomenon is similar to the one observed in
\DMW\ which is possibly an indication of some
`phase transition' at the boundary.

We can also contemplate more complicated possibilities.
For example, if $y_1$ and $y_2$ are two fixed points
of $R$ that are not related by $S$, then we can place
eight 5-branes at $y_1$ and
eight at $y_2$. The remaining $16$ branes have to
be placed at the images of these two points under $S$.
In this case one would obtain $U(4)\times U(4)$
gauge group with two copies of $(4, {\bar 4})$ from
each of the fixed points. Now the anti-self-dual
tensors coming from twisted sectors at both $y_1$ and $y_2$
will be needed for anomaly cancellation.

\newsec{Discussion}

We have constructed a string theory that does not
seem to be connected to the known string vacua because
we have a different number of tensor multiplets.
It cannot be viewed  as a compactification of Type-I theory
because the orientifold symmetry mixes nontrivially with
internal symmetries of the $K3$. We have discussed here
only the simplest example but quite clearly there is
a whole class of models one can consider at different
points in this moduli space. Work on some of these models
is in progress and will be reported elsewhere \toappear.
Models with multiple tensor multiplets have been considered before in
\refs{\SagnI, \SagnII} although from a somewhat different
point of view.

By analogy with \Stro\ one can ask if
these theories are connected to other theories by a
phase transition.
In six dimensions, infrared dynamics
is trivial, so it would seem
impossible to change the number of anti-self-dual tensors
because one can simply count the states in the infrared.
Such a transition can occur only if there is non-trivial
infrared dynamics at special points in the moduli space
analogous to the situation considered in \WittII.
Perhaps the boundary in the tensor-multiplet
moduli space where the kinetic term for the gauge fields changes
sign is related to such a phase transition.

Finally, one can ask about the duals of the theories that
we have constructed. A.~Sen has informed us
that at a generic
point in the moduli space with $U(1)^8$ gauge symmetry, one can obtain
identical spectrum by
considering an orbifold of M-theory compactified on $K3 \times S^1$
\Sen. In this theory the vector mulitplets arise
from the untwisted sector whereas the tensor multiplets arise from the
addition of 5-branes of M-theory by a reasoning similar to
\refs{\DaMu, \Witt}.
This is complementary to our construction where the tensor multiplets
arise from the untwisted sector (on a smooth $K3$) and the vector
multiplets arise from the addition of 5-branes.
In a recent paper that appeared after this work was completed,
C.~Vafa has obtained
identical spectrum by a compactification of `F-theory' \Vafa.
It is plausible that these three models can be related to one another
by duality.

\bigskip
\leftline{ \secfont Acknowledgements}
\bigskip
We are grateful to J.~Polchinski,
J.~H.~Schwarz, A.~Sen, and E.~Witten for valuable discussions.
This work was supported in part by the U. S. Department of Energy
under Grant No. DE-FG03-92-ER40701.
\vfill
\eject
\listrefs
\end